\documentclass[12pt]{article}
\usepackage{amsmath,amsthm, amssymb}
\usepackage{graphicx}
\usepackage{leqno}
\usepackage{hyperref}
\numberwithin{equation}{section}
\begin{document}
\title{\vskip-40pt Causality, Measurement, and Elementary  Interactions}
\author{Edward J. Gillis\footnote{email: gillise@provide.net}}
				
\maketitle

\begin{abstract} 

\noindent 

Signal causality, the prohibition of superluminal information transmission, is 
the fundamental property shared by quantum measurement theory and relativity, 
and it is the key to understanding the connection between nonlocal measurement 
effects and elementary interactions. To prevent those effects from transmitting 
information between the generating and observing process, they must be induced 
by the kinds of entangling interactions that constitute measurements, as implied 
in the Projection Postulate. They must also be nondeterministic as reflected in 
the Born Probability Rule. The nondeterminism of entanglement-generating 
processes explains why the relevant types of information cannot be instantiated 
in elementary systems, and why the sequencing of nonlocal effects is, in 
principle, unobservable.  This perspective suggests a simple hypothesis about 
nonlocal transfers of amplitude during entangling interactions, which yields 
straightforward experimental consequences.

\end{abstract}

\section{Introduction}

Measurements are carefully designed arrangements of elementary, correlating 
interactions, but standard formulations of quantum theory treat the act of 
measurement as irreducible. This explanatory gap between microphysical 
evolution and macroscopic outcomes is a serious problem for the conventional 
theory of measurement. However, it should not lead us to overlook some 
particularly elegant features of that theory, and the extent to which it 
permeates and shapes fundamental physics.

The central place of the measurement postulates in contemporary physics is 
illustrated in the following passage from one of the most widely used textbooks 
on quantum field theory\cite{Peskin}. After showing that spacelike propagation 
is an inevitable consequence of the relativistic field equations, the authors 
say (p. 28): \begin{quotation}\noindent "To really discuss causality, however, 
we should ask not whether particles can propagate over spacelike intervals, but 
whether a \textit{measurement} performed at one point can affect a measurement 
at another point whose separation is spacelike."\end{quotation} This passage 
clearly recognizes that, at the observational level, the standard projection 
and probability postulates constitute a completely causal and relativistic 
account. It also shows that the characterization of quantum field theory as 
incorporating the principle of causality at the most elementary level depends 
critically on the connection to macroscopic measurement. This critical 
dependence stems from the nonlocal
effects that are inherent to quantum theory. 

The reality of nonlocal projection effects was demonstrated by 
Bell\cite{Bell_1,Bell_2}, and confirmed experimentally by 
Aspect\cite{Aspect_1,Aspect_2}. Bell showed that the correlations exhibited 
between the results of distant measurements on pairs of entangled particles 
could not be produced unless one of the measurements acts across spacelike 
intervals to affect\footnote{Bell's demonstration does not contradict the 
statement of Peskin and Schroeder since they are referring to effects on 
the \textit{total} probability of an outcome, while Bell's correlations describe 
\textit{conditional} probabilities of an outcome given specific results of 
distant 
measurements.} the other measurement. 

The recognition of real nonlocal actions has led contemporary physics to 
modify the traditional concept of causality which implies strictly local 
propagation. The modern rendering of the principle as a prohibition of 
superluminal signaling reflects a significant generalization of the 
classical notion. This perspective allows nonlocal effects, 
provided that they cannot transmit information. 

The differences between the classical and contemporary notions were discussed 
by Bell in his last essay\cite{Bell_LNC}. His main arguments have been further 
elucidated by Maudlin\cite{Maudlin} and Norsen\cite{Norsen_1}. Local causality 
captures our intuitive notion of physical processes propagating continuously 
through space, and is sufficient to insure that those processes can be described 
in a Lorentz covariant manner. Signal causality fails in both these respects. It 
is largely our inabilty to consistently describe projection as a process at all, 
or accomodate it readily within a relativistic spacetime that has made it difficult 
to construct a coherent microphysical explanation of measurement effects. This 
apparent obstacle to a fundamental explanation of measurement is what led Bell 
to express his distaste for the contemporary viewpoint. 

Despite these legitimate concerns, signal causality, when construed as 
a relationship among \textit{purely physical} processes and systems, is extremely 
useful as a guiding principle in the formulation of fundamental laws. 
Svetlichny\cite{Svetlichny_1,Svetlichny_3} has emphasized this point. He argues 
that superluminal communication is  a \textit{generic} property of physical 
theories. So those that exhibit strong nonlocal quantum correlations without 
signaling are very special. In Svetlichny's words, this property imposes 
a rigidity on fundamental theories. This rigidity means that the principle of 
causality is sufficient to determine many of the essential characteristics of 
the theories that embody it. 

Note first that it is signal causality that has guided the development of quantum 
field theories and shaped some of their most fundamental properties. As a further 
illustration of this point consider the Born Probability Rule\cite{Born}. Given 
the nonlocal projection that results from measurement, the no-superluminal-signaling 
principle determines precisely how probabilities must be assigned to possible 
outcomes. (This is demonstrated in Gleason's theorem\cite{Gleason} since the 
no-superluminal-signaling requirement entails Gleason's crucial assumption of 
noncontextuality.) The rule is so familiar that we take it for granted, but as 
Bell points out\cite{Bell_prob} it is quite easy to imagine other reasonable 
prescriptions for ascribing probability. For example, one could make it 
proportional to the absolute value of the amplitude rather than the square of 
the amplitude. One could associate amplitudes with cosines and make the 
probability some function of the angle. Equal probabilities could be assigned 
to all nonzero amplitudes, or one could dictate a deterministic outcome based 
on the largest amplitude. The point is that \textit{any} of these reasonable 
alternatives would enable transmission of information across spacelike intervals. 
The no-signaling principle selects the Born Rule as the \textit{only} way to 
assign probabilities to outcomes. 

Although the Born Rule is very specific, it does illustrate a very general, 
deep feature of causal, nonlocal theories, viz., indeterminism. The generality 
of the connection between nondeterminism and causality in the presence of 
nonlocal effects was brought into prominence over the last couple of 
decades through the efforts of a number of resarchers. Elitzur\cite{Elitzur_92} 
argued that indeterminism should be taken as a fundamental principle of quantum 
theory that is deeply connected to both the prohibition of superluminal 
information transmission and the Second Law of Thermodynamics\footnote{The 
possibility of a connection between measurement effects and entropy increase 
was also noted by von Neumann\cite{von_Neumann,von_Neumann_2}. The same work is also usually 
recognized as giving the first formulation of the Projection Postulate.}. 
Popescu and Rohrlich\cite{Popescu,P_R_2,P_R_3}, building on earlier suggestions 
by Aharonov\cite{Ahar_Rohrlich} and Shimony\cite{Shimony_1,Shimony_2}, proposed 
taking nonlocality and relativistic causality as axioms, and deriving 
indeterminism as a consequence. Additional arguments that nonlocal effects must 
be nondeterministic \textit{with respect to observable quantities} in order to 
prevent superluminal information transfer have been given in several 
works\cite{Svetlichny_1,Svetlichny_3,Svetlichny_2,Elitzur,Elitzur_2,Masanes,Qi-Ren,Valentini}. 
The general relationship of causality, relativity, and observability to 
indeterminism has been eloquently summarized by Elitzur and Dolev\cite{Elitzur}. 
They note that  "\textit{Hidden variables must be forever-hidden variables}". This 
view is reinforced by the work of Valentini\cite{Valentini} which shows that the 
prevention of superluminal signaling in the de Broglie-Bohm\cite{deBroglie,Bohm} 
theory depends on the initial distribution of particles being consistent with the 
initial wave function. Other types of initial  distributions predict experimental 
differences between Bohmian theory and standard quantum mechanics. The critical 
point is that these differences, if observed, would constitute superluminal signals.

Aharonov has described this situation as follows\cite{Rohrlich}:
\begin{quotation}
        ``Instead of asserting, with Einstein, that 'God does not play dice!' 
        we should ask ourselves, 'Why does God play dice?' i.e., What new 
        possibilities does a non-deterministic universe offer? A first answer 
        would be, non-determinism allows a universe that is self-consistent, 
        and causal in the relativistic sense, to be nonlocal.''	\end{quotation}     
The picture that emerges is that the nonlocality of the Projection Postulate is 
tempered by the limits on determinism implied by the Born Rule to maintain causality. 

By recognizing the limits on determinism as an essential feature of contemporary 
theory we solve one of the basic riddles of causality: How can real changes in 
definite physical states propagate across spacelike intervals without the 
transmission of information? The answer is that the limits on determinism 
imply that the concept of information is just not definable at the most 
elementary scales. To say that information about a physical state exists in 
a particular system implies that the state can be captured and reliably 
replicated. We know from the no-cloning theorem\cite{No_Clone} that, in 
general, this cannot be done for elementary systems. It does not make much 
sense to say that an isolated elementary particle carries information if the 
laws of physics forbid the extraction of that information\footnote{Verification 
measurements that register previously prepared states do not serve as 
counterexamples to this point. Particles in prepared states are not isolated; 
information is instantiated in their relationship to the preparation apparatus 
and process.}. The underlying reason that one cannot find out everything about 
arbitrary states is that the physical processes that constitute measurements 
are partially \textit{nondeterministic}. The concept of information requires 
a degree of determinism that, in general, just does not exist in interactions 
involving a few elementary particles. It takes a large number of such interactions 
to insure that an outcome of a probabilistic process is fully resolved. The 
relevant properties must also be reflected in the states of enough individual 
particles so that the record of the outcome is not wiped out by further 
probabilistic occurrences. Information can only be said to meaningfully exist 
at a scale at which statistical stability is reasonably assured.\footnote{The 
situation is somewhat analogous to the relationship between statistical mechanics 
and thermodynamics where the concepts of temperature, pressure, and entropy are 
defined only for systems that are large enough to approximate thermal equilibrium.} 

So the no-superliminal-signaling principle entails the Born Rule (given the 
vector space structure of quantum theory and nonlocal projection effects). 
This precludes violations of macroscopic causality, and the indeterminism 
that is implied also explains how real physical changes can propagate across 
spacelike intervals without the transmission of information. The indeterminism 
prevents information from being instantiated in isolated elementary systems.

But there is at least one further implication of signal causality that provides 
a critical clue in constructing a microphysical explanation for nonlocal measurement 
effects. As already emphasized, to serve as a useful guide, this principle must 
be interpreted as a relationship among \textit{physical} entities without any 
reference to intelligent observers. Therefore, any physical process that can 
observe the results of nonlocal effects such as wave function collapse must, 
itself, be capable of inducing those effects. The physical processes that bring 
about collapse are measurements. If measurements did not themselves bring about 
the nonlocal effects, they would be observing the results of \textit{other} 
physical processes that occur at spacelike separation. Aside from any question 
about whether intelligent observers are signaling one another, information about 
the occurrence of those other processes would be transmitted across spacelike 
intervals, violating causality. To avoid such violations, one must assume that 
it is something \textit{intrinsic to the measurement process} that induces the 
nonlocal quantum effects that are observed. 

The connection between measurement and projection, is, of course, explicitly 
spelled out in the Projection Postulate, but the fact that it is a critical 
implication of causality is often overlooked. To maintain a meaningful notion 
of causality at the microphysical level we should assume that nonlocal collapse 
effects result from the interactions that establish correlations between 
elementary particles. These are what measurements are made of. When these 
interactions involve part, but not all, of the of the subject wave functions, 
they can generate entanglement and the problematic, nonlocal effects associated 
with it. In the hypotheses offered here, these  \textit{elementary entangling 
interactions} will be responsible both for creating entanglement relations 
and for the projection-like effects that eventually break them.

For the proposed explanation to make sense, all that we need to assume is that 
entanglement, the generation of entanglement, and the nonlocal correlations 
implied by it are genuine physical phenomena. The reality of these effects has 
been well established, both theoretically and experimentally, and they constitute 
the truly distinctive characteristics of quantum theory. This approach allows us 
to get to the heart of the problem without getting hung up on specific aspects 
of the formalism that might eventually be superseded. We will also see that this 
assumption solves the so-called "basis selection problem" in a very simple and 
natural way. 

The account of nonlocal collapse effects that I will offer turns on a willingness 
to acknowledge (at least provisionally) that current theory assumes that there 
are \textit{essential} limits to determinism, but these limits do \textit{not} 
call into question the reality of elementary processes. The need to distinguish 
between physical realism and determinism is eloquently described by Bradley\cite{Bradley}. 
After expressing a generally positive view of determinism he says: 
\begin{quotation}
            "Nevertheless, I have to admit that there is no a priori reason 
             why either a metaphysical or scientific realist should be a determinist. 
             Neither form of realism actually entails determinism. Perhaps God 
             does play dice with the cosmos after all. That the universe should 
             be indeterministic - that events at the microphysical level, in 
             particular, should be uncaused - is entirely conceivable." 	\end{quotation} 
Gisin\cite{Gisin_1,Gisin_2} has also discussed the need to keep the distinction 
clear, and pointed out some of the confusions that arise when these two different 
concepts are conflated. For physicists, it is reasonable to take realism (a belief 
in the objective existence of external reality) as an axiom, but determinism 
should be treated as an hypothesis. Hume\cite{Hume} made this point over 
250 years ago when he pointed out that there are no necessary connections 
in experience. 

To summarize, given the real nonlocal effects pointed out by Bell's analysis, 
the principle of signal causality entails that the processes that observe these 
effects must be the ones that induce them, and that they must be nondeterministic 
with probabilities in accord with the Born Rule. From this perspective, the 
standard measurement postulates are straightforward consequences of the prohibition 
on superluminal information transmission. When we translate these implications to 
a more fundamental level we see that the nondeterministic aspects of elementary 
interactions prevent information from being instantiated in very small physical 
systems, and that the nonlocal effects must be induced by elementary entangling 
interactions. 

These inferences will be exploited in Section 3 to develop a microphysical account 
of nonlocal measurement effects. But first we must try to reconcile our understanding 
of relativity and spacetime structure with such nonlocal effects.

\section{Causality, Relativity, and the \newline Sequencing of Nonlocal Effects}

Causality and relativity are intimately related. In elementary discussions, 
this close relationship is usually attributed to spacetime structure, and it 
is easy to see how both properties are maintained when physical processes 
are confined to the light cone. Above, we saw that, with nonlocal measurement 
effects, the preservation of causality entails the Born Rule. Given the close 
connection between the two principles, we should expect that relativity 
also depends on the rule. This dependence can be illustrated with a simple 
two-particle entangled system. 

Consider a situation with identical particles in the state: 
 $ \alpha|x_1\rangle|x_2\rangle +  \beta|y_1\rangle|y_2\rangle$, 
 with $ |x\rangle$ and $|y\rangle$ orthogonal, and 
 $\alpha\alpha^* + \beta\beta^* = 1$.	Define an alternate basis for the 
 second particle:
\[  |u_2\rangle =  \gamma|x_2\rangle + \delta|y_2\rangle, \; |v_2\rangle =  \delta^*|x_2\rangle - \gamma^*|y_2\rangle.\]
The original basis can be expressed as:
\[ |x_2\rangle = \gamma^*|u_2\rangle + \delta|v_2\rangle , \; |y_2\rangle =  \delta^*|u_2\rangle - \gamma|v_2\rangle.  \]
One can represent the first particle in the $|x\rangle,|y\rangle$ basis, 
and the second particle in the $|u\rangle,|v\rangle$ basis: 
\[   \alpha|x_1\rangle (\gamma^*|u_2\rangle + \delta|v_2\rangle)  + \beta|y_1\rangle (\delta^*|u_2\rangle - \gamma|v_2\rangle) \]
\[  = \; \alpha \gamma^*|x_1\rangle|u_2\rangle + \alpha \delta|x_1\rangle|v_2\rangle + \beta \delta^*|y_1\rangle|u_2\rangle - \beta\gamma|y_1\rangle|v_2\rangle. \]
In Bell-EPR\cite{Bell_1,Bell_2,EPR}-type experiments one can measure either 
particle in any basis whatsoever. One can also measure one of the particles 
without measuring the other. The no-superluminal-signaling principle implies 
that a measurement on one particle must not affect the outcome probabilities 
of measurements on the other. Therefore, the sum  of probabilities conditioned 
on specific outcomes of a measurement must equal the total probability when 
no measurement is made on the other particle. Specifically, for a $|u\rangle$ 
outcome on the second particle they must satisfy:
		\[P(\alpha)P(\gamma^*) + P(\beta)P(\delta^*) = P(\alpha \gamma^*) + P(\beta \delta^*),\]  	
where $P(\alpha)$ denotes the probability associated with the amplitude, $\alpha$. 
This condition is obviously fulfilled by the Born Rule 
($\alpha\alpha^*\gamma\gamma^* +  \beta\beta^*\delta\delta^*$ ), and it is easy 
to show that this is the \textit{only} way to satisfy it.\footnote{See Gleason's 
theorem\cite{Gleason}.} This equality has been derived from the assumption of 
\textit{causality}, i.e., that the total outcome probability of a measurement on 
particle 2 is not affected by a measurement on particle 1. However, one can also 
take it as a statement that the sequence of measurements can be freely 
interchanged, since the right-side expression implies that the measurement on 
the second particle is made first, and the left side can be read in either order. 
This freedom to interchange the sequence of spacelike-separated events is a hallmark 
of \textit{relativity}. So, relativity, like causality, is \textit{not} simply 
a consequence of spacetime structure. In describing nonlocal measurement effects, 
it is dependent on the specific form of indeterminism embodied in the Born 
Rule. A very similar point is made by Elitzur and Dolev\cite{Elitzur}. 

Relativity is a property of our \textit{theories}. It is rooted in several deep 
features of those theories - not just space-time relationships. These features 
include fundamental limits on observability that are tied to nondeterministic 
effects of elementary interactions. 

We have seen that causality requires that nonlocal measurement effects must be 
probabilistic. This nondeteminism, in turn, implies that the effects are 
irreversible.\footnote{The connection between indeterminism and irreversibility 
has been pointed out in several places by Elitzur\cite{Elitzur_92}, and by 
Elitzur and Dolev\cite{Elitzur,Elitzur_2}.} It is extremely difficult to construct 
a coherent account of such irreversible effects without assuming that they occur 
in some definite sequence.\footnote{Maudlin\cite{Maudlin} has pointed out many 
of the complications for such attempts.} I will shortly offer additional arguments 
that we should attribute some objective, though unobservable, sequence to these 
nonlocal effects. 

Such objective sequences of spacelike-separated events are, of course, at odds with 
our ideas about relativity. This apparent conflict can be resolved in essentially 
the same way in which one reconciles the propagation of real nonlocal effects with 
the impossibility of superluminal information transmission. The effects of the 
sequencing of nonlocal actions are not observable because  those actions occur at 
a level well below that at which the relevant information could be instantiated. 
Observations are macroscopic, or at least, mesoscopic processes that involve the 
recording of reasonably definite outcomes of a series of probabilistic events. Any 
single observation is consistent with a wide range of sequences of nondeterministic 
microphysical events. It is the many-to-one map of \textit{possible} sequences to 
a particular observed outcome that makes a relativistic description appropriate 
for situations with multiple spacelike-separated measurements. 

So, in measurement situations, relativistic descriptions do not capture all 
space-time relationships because some of those relationships leave no physical 
trace. Our ability to apply such descriptions depends, in large part, on in-principle 
limitations to the amount of information generated by elementary processes.

The idea that relativistic symmetries reflect limits on information is not 
necessarily tied to a fundamental indeterminism.  A proposal by  't Hooft\cite{Hooft_2} 
is based on the idea that there is an essential information loss mechanism between 
the most basic ontological level and the level of quantum description: 
   \begin{quotation}
              ``We ... argue that symmetries such as rotation ..., translation, 
                Lorentz invariance, ... gauge symmetries and coordinate 
                reparametrization invariance, might all be emergent... 
                [and] that information is not conserved in the deterministic 
                description.'' 	\end{quotation} 

While deterministic extensions of probabilistic theories are always possible, in 
principle,\footnote{ One can always "fill out" the predictions of a probabilistic 
theory by providing a (possibly infinite) list that determines everything that was 
initially left undetermined. Interesting deterministic extensions generate the list 
dynamically from initial conditions.} the goal here is to outline a logically 
coherent microphysical explanation of measurement effects \textit{within} the 
framework of contemporary physics. Causality and relativity are defining features of 
that framework. Their preservation requires that we accept the Born limit on 
determinism as a guiding principle. This perspective allows us to view causality 
and relativity as consequences of fundamental physical law, even though they do 
not derive only from the structure of spacetime.

The need to invoke the limits on determinism and information in order to save 
relativity stems from the fact that there is an implied \textit{sequencing} of the 
nonlocal effects described by the measurement postulates. Although nonlocal 
propagation, in itself, can present challenges to the construction of a relativistic 
description of physical processes based solely on the structure of spacetime, these 
challenges are not necessarily insurmountable. Some aspects of nonlocality can be 
dealt with, as indicated in the passage from Peskin and Schroeder\cite{Peskin} 
quoted earlier, and as discussed at some length by Maudlin\cite{Maudlin}. However, 
attribution of objective sequences to large sets of spacelike-separated events 
creates extreme difficulties for the notion that relativity concerns only spacetime 
relationships. Let us now look at some of the reasons that one would want to posit 
such sequences, even though they force us to alter some of our ideas about the 
basis for relativity.\footnote{The need to assume some sort of sequencing in 
any logically coherent account of measurement effects has been noted  
frequently in the literature\cite{Albert,Maudlin}}

I will first look at the issue of sequencing as it applies to macroscopic 
measurements, and then extend the analysis to the elementary interactions that 
constitute those measurements. Recall the simple entangled state described above. 
The state can be represented in two ways depending on which measurement one 
wants to emphasize.
\[  \begin{array}
{ll} & \alpha|x_1\rangle \, \otimes\, (\gamma^*|u_2\rangle + \delta|v_2\rangle)  
\quad + \quad \beta|y_1\rangle \, \otimes \, (\delta^*|u_2\rangle - \gamma|v_2\rangle)  \\
 =  & (\kappa|x_1\rangle| + \lambda |y_1\rangle) \, \otimes \, \sigma|u_2\rangle  
\quad + \quad (\mu|x_1\rangle| + \nu |y_1\rangle) \, \otimes \, \tau|v_2\rangle . 
\end{array} \] 
The relevant relationships between the two sets of 
coefficients are: $\alpha\gamma^* = \kappa\sigma$, \newline
$\beta\delta^* = \lambda\sigma$, $\alpha\delta = \mu\tau$, and $\beta\gamma = -\nu\tau$.

Suppose that measurements on the two particles are \textit{timelike} separated. 
A measurement on particle 1 yields an $|x_1\rangle$ result, and a later one on 
particle 2 produces a $|u_2\rangle$ outcome. The sequence of projections is: 
\[  \begin{array}
{ll} & \alpha|x_1\rangle \, \otimes\, (\gamma^*|u_2\rangle + \delta|v_2\rangle)  
\quad + \quad \beta|y_1\rangle \, \otimes \, (\delta^*|u_2\rangle - \gamma|v_2\rangle)  \\
& {\Longrightarrow} \qquad |x_1\rangle\otimes(\gamma^*|u_2\rangle + \delta|v_2\rangle)
   \qquad {\Longrightarrow}\qquad |x_1\rangle|u_2\rangle.    
\end{array} \] 
If the order of the measurements were reversed the projection sequence would be:
\[  \begin{array}
{ll} & (\kappa|x_1\rangle + \lambda |y_1\rangle) \, \otimes \, \sigma|u_2\rangle  
\quad + \quad (\mu|x_1\rangle + \nu |y_1\rangle) \, \otimes \, \tau|v_2\rangle \\
& {\Longrightarrow} \qquad (\kappa|x_1\rangle+ \lambda|y_1\rangle) \, \otimes \, |u_2\rangle
   \qquad {\Longrightarrow}\qquad |x_1\rangle|u_2\rangle.    
\end{array} \] 
The correlations between the outcomes are encoded in the joint probabilities. 
The sequential projections provide a very simple way to understand these 
correlations. Because $P(\alpha)*P(\gamma^* )= P(\kappa)*P(\sigma$) they are 
not affected by reversing the order of the measurements.  

In these timelike-separated cases we can apply the Projection Postulate in 
a completely straightforward manner. The first measurement collapses the wave 
function to the observed outcome \textit{and to the state of the other particle 
that is correlated with that outcome.} The second measurement then acts on the 
state to which the system has collapsed. The measurements act (possibly nonlocally) 
on the state by interacting locally with one of the branches of the wave function. 
Nonlocal effects are mediated by entanglement relations. The nonlocal effects are 
at odds with our classical intuitions about causality, but the logical relationships 
are quite simple. 

We can retain this straightforward picture for situations in which measurements 
are spacelike-separated if we are willing to attribute \textit{some} sequence 
to those measurements. The Bell-EPR correlations between outcomes are the same 
as those just described for the timelike-separated measurements. As just shown, 
the joint probabilities of outcomes (and, hence, the correlations) are not 
affected by which measurement is sequenced first. The Born Rule prevents any 
observable superluminal effect of one measurement on the other, and it also 
precludes any detection of what the sequence is. Note that, with the assumption 
of sequencing, \textit{the only nonlocal action is by the measurement on the state}. 
The measurements do not need to cooperate, conspire, or communicate with each 
other across a spacelike interval in order to maintain the Bell correlations. 

This last point is relevant because the goal here is to outline a logically simple 
microphysical explanation of wave function collapse, viewed as a real physical 
process. That explanation will consist of two assumptions. One of these concerns the 
sequencing of spacelike-separated elementary, entangling interactions; the other 
concerns the nondeterministic, nonlocal action of those interactions on the wave 
functions of the interacting particles. One might think that, in the interest of 
preserving more of the relativistic spacetime structure, it would be desirable to 
dispense with the sequencing assumption, and posit some additional type of nonlocal 
effect that enables spacelike-separated measurement processes to cooperate. 
However, it is very difficult to develop any explanation that does not make 
some analogous assumption about the ordering of events in spacetime that is 
seriously at odds with conventional ideas about relativity.

In situations with just two well-defined, localized, spacelike-separated 
measurements, it is tempting to view them as acting jointly in the reference 
frame in which they are simultaneous, to produce a single projection to the 
final state:
\[  \begin{array}
{ll} & \alpha \gamma^*|x_1\rangle|u_2\rangle + \alpha \delta|x_1\rangle|v_2\rangle 
  + \beta \delta^*|y_1\rangle|u_2\rangle - \beta\gamma|y_1\rangle|v_2\rangle
   \qquad {\Longrightarrow}\qquad |x_1\rangle|u_2\rangle.    
\end{array} \]  However, if more than two processes are involved, such a frame 
does not always exist. Formally, one can still view the sets of spacelike-separated 
interactions as constituting a single measurement process.\footnote{Thanks to the 
reviewer for emphasizing this point.} But the problem becomes more complicated when 
we consider mixed sets of interactions, involving both spacelike and timelike 
separations between them. Although they can still be viewed, from a formal point 
of view, as one measurement process, this would take us even farther from an 
understanding of the nonlocal actions \textit{as real, physical effects}. 

To see this, suppose that a set, A, of interactions is timelike earlier than B, 
and that both A and B are spacelike-separated from C. Considered by itself, we 
would expect A to yield a definite outcome, as a result of some physical process. 
To suggest that this outcome must be postponed until the ocurrence of B (and 
possibly even later events) because of something that might be happening at 
a spacelike separation is counterintuitive, and makes any potential physical 
explanation much more convoluted. It also raises the possibility of indefinite 
further postponements. (A similar point about indefinite regress is made by 
Maudlin\cite{Maudlin} concerning Cramer's transactional 
interpretation\cite{Cramer}.) To try to associate timelike-separated processes 
with distinct projections assumes some criteria for grouping processes across 
spacelike intervals: should C be bracketed with A or with B? Such grouping 
criteria would conflict with relativistic structure just as much as the assumption 
of sequencing of spacelike-separated events.

The motivation to assume some type of sequencing becomes even stronger when one 
contemplates possible explanations of projection in terms of elementary 
interactions. Any real, measurement process consists of myriad elementary correlating 
interactions, with both spacelike and timelike relations to one another. So, even 
in the case of a single, reasonably well-defined macroscopic measurement, we would 
face all of the issues described in the previous paragraph. One should also note 
that entanglement relations are generated by exactly these kinds of interactions, 
and that \textit{these entanglement relations both mediate and define wave function 
collapse}. 

This last point implies that the generation of entanglement, itself, is an 
important type of nonlocal action. Although the entangling effects are 
deterministic, they define the scope of the collapse effects which are not. 
At a fundamental level, the entanglement relations determine exactly which 
elementary systems are directly affected by the probabilistic projections.  
The fact that causality entails that the projection effects are tied to 
the entangling interactions helps to simplify the overall picture. But it 
also means that even simultaneous interactions involving the same entangled 
system must be sequenced.

To make clear just what is being proposed, let us compare the hypothesized 
sequencing to the de Broglie - Bohm theory\cite{deBroglie,Bohm}. Bohmian 
mechanics assumes an absloute time (at least in the most straightforward 
version). This is obviously at odds with the idea of a relativistic spacetime, 
and it implies much stronger temporal ordering relations. However, it still 
allows large numbers of entangling interactions to occur simultaneously. 
The proposal outlined in the next section associates these individual 
elementary interactions with the collapse effects. Because these effects are 
probabilistic, and, hence, irreversible, logical simplicity strongly 
suggests that even simultaneous interactions should be assigned some objective 
(though indiscernible) sequence. The reason that Bohmian theory is able to 
present a logically coherent account of measurement outcomes is that, in it, 
the wave function \textit{never collapses}. The outcomes are determined by the 
action of the wave function in conjunction with the simultaneous positions of 
all of the particles involved. Portions of the wave function become 
irrelevant\footnote{Because they are so far removed from the system location 
in configuration space.}, but they never vanish.  

The sequencing of spacelike-separated entangling interactions that is 
proposed here goes beyond what is implied by a notion of absolute time.  
For the reasons described earlier, we want to insure that entanglement 
relations are well defined at the time and place of each individual 
interaction, and we also want to avoid the need for interactions to 
cooperate across spacelike intervals. Therefore we will assume that, 
no matter what spacelike hypersurface is given, there is some objective 
(though unobservable) sequence of the nonlocal effects of interactions 
on that surface that involve the same entangled system. So if A and B are 
spacelike-separated interactions involving particles that are entangled 
with one another, either A is ``prior'' to B, or B is ``prior'' to A.

There are some similarities between this proposal and Bohmian theory. 
One can describe the sequencing by reference to a set of preferred 
(unobservable) spacelike hypersurfaces. These need not be hyperplanes, 
as they are in the de Broglie - Bohm account. However, a more critical 
difference is that \textit{the set of hypersurfaces does not form 
a foliation of spacetime}, because successive surfaces will not be 
completely disjoint. To describe the idea more precisely, let us 
consider two spacelike-separated interactions, 
A and B, that involve the same entangled system. Imagine that the 
evolving hypersurface lies immediately earlier than both A and B. If 
the nonlocal effects of interaction A are to be sequenced prior to those 
of B, then the surface should be pushed forward in the local vicinity of A, 
but held back in the region of B. This situation can be characterized with 
a parameter, $s$. Let $s_{0}$ label the hypersurface that was just 
described, and let $s_{1}$ label the surface that is pushed forward 
to include A, but remains stationary in the neighborhood of B (and, 
hence, is still earlier than B). The surface that is eventually 
pushed forward to include B can be labeled $s_{2}$. The sequencing 
of A prior to B is reflected in the fact that $s_{1} < s_{2}$ (since 
$s_{1}$ labels the first surface on which A lies, and $s_{2}$ labels 
the first surface to include B). \footnote{We can fill out the description 
by using local time variables $d\tau_{A}$ and $d\tau_{B}$. Since A is 
sequenced first, we get  $d\tau_{A}/ds > 0, d\tau_{B}/ds = 0$ during the 
interval $s_{0} < s < s_{1}$.} 

This approach is closely akin to the many-time formalism of quantum 
theory\cite{Tomonaga,Schwinger}. This formalism has been exploited by 
Bell\cite{Bell_jump}, by Bohm and Hiley\cite{Bohm_Hiley}, and by 
Bedingham\cite{Bedingham} to deal with the issues discussed here. 
However, it is important to (again) note that we do not revert 
to a foliation of spacetime for simplicity, as is sometimes done. 
The idea is to exploit the full freedom allowed by the multiple-time 
approach in order to allow for a more complete sequencing of the 
relevant interactions.

This formulation enables one to describe the nonlocal effects of each 
interaction involving the same entangled system individually, in sequence. 
This allows us to construct a logically simple account of wave function 
collapse. On any particular surface, entanglement relations and states are 
well defined. The nonlocal effects are induced by the interaction on the 
state, and they propagate along the surface. The specific nature of the 
effects (to be described in the next section) reproduces the standard quantum 
probabilities and correlations, and it insures that sequencing of 
spacelike-separated interactions is unobservable. With respect to a standard 
foliation of spacetime, the family of surfaces hypothesized here is 
\textit{overcomplete}. Any particular event can lie on many such surfaces.

This description of nonlocal collapse effects is \textit{not} Lorentz-invariant 
or covariant. The idea is that a standard relativistic account can be recovered 
when one averages over all of the possible families of evolving surfaces (and 
sequences) that are consistent with the observed outcome. This averaging is 
roughly similar to a statistical mechanics description that averages over many 
possible microconfigurations.

The evolving surface described here is intended to be relevant only for the
description of nonlocal collapse effects. Ordinary deterministic processes are 
constrained to propagate within the light cone and are not affected by this 
additional spacetime structure.  

The evolution of the surface is constrained by entanglement relations in order to 
insure that interactions affecting a given system are sequenced one at a time, 
and by the requirement that it must remain spacelike. Other than this, the 
evolution is assumed to be random. So, just as in conventional formulations of 
relativity, there is no preferred reference frame. 

I will not attempt to completely fill out this picture here. Instead, I will 
consider \textit{discrete} sets of interactions. An hypothesis concerning the 
nonlocal effects of such interactions is presented in the next section. It will 
reproduce the Projection Postulate and the Born Probability Rule at the macroscopic 
level, and it will entail (due to the indeterminism of the nonlocal effects), that 
every possible sequence of spacelike-separated interactions is consistent with the 
observed outcome.  The assumption that there is an objective sequence of these 
interactions provides logical coherence to the account. The fact that the sequences 
are unobservable maintains consistency with relativity.

\section{Connecting Measurement to \newline Elementary Interactions}

Measurements are essentially detections, or failures to detect. Ultimately, they 
register the presence or absence of particles.The state vector collapse that is 
observed implies the transfer of the wave function's amplitude either completely 
into or out of the region of the interactions that constitute the measurement. 
Cases in which the correlating interactions involve the complete wave function of 
the measured system correspond to simple verification measurements, i.e., those in 
which the system is already in an eigenstate of the measured observable. It is when 
some of the relevant interactions involve part, but not all, of the subject wave 
function, that the problematic, nonlocal transfers of amplitude can 
occur.\footnote{These ``partial'' encounters are one of the principal types of 
interactions that generate entanglement. Processes such as parametric down 
conversion in which particles are \textit{created} in an already entangled state 
are the other principal type.} 

It is the interactions that also \textit{define} the measurement basis. At 
a macroscopic level the interaction Hamiltonian of the measurement apparatus 
determines what is being measured. This has been clearly shown by Laura 
and Vanni\cite{Interact_Basis}. They demonstrate that different Hamiltonians 
measure different observables.\footnote{These arguments are related to earlier ones 
developed by Zurek\cite{Zurek1}, in his description of  the decoherence approach.}

At an elementary level, the interactions determine the basis simply by defining 
a bifurcation of the wave function into interacting and noninteracting parts. In 
this respect the individual interactions exhibit a clear parallel to the ``yes-no'' 
detection-like character of macroscopic measurements. This close parallel suggests 
a very straightforward hypothesis about how elementary entangling interactions 
eventually bring about wave function collapse. Since collapse is equivalent to 
the complete transfer of amplitude either into or out of the measurement region, 
it is natural to suppose that each individual interaction transfers some small 
amount of amplitude either into or out of the state resulting from that interaction. 
With enough such shifts, eventually all of the amplitude will be transferred. 
(Since different interactions can split the wave function in different ways, 
macroscopic measurements can have arbitrarily many possible outcomes.) Note that 
it is the \textit{linearity} of the Schr\"{o}dinger equation that permits nonlocal 
shifts in amplitude without otherwise disturbing the ordinary deterministic 
evolution of the system.

Of course, the shifts must reproduce the Born Rule at the level of macroscopic 
outcomes, but this is not difficult to arrange. According to the rule, the 
probability of an outcome is equal to the absolute square of the amplitude of 
the relevant wave function component, so it is convenient to give this quantity 
a name\footnote{Calling it a ``probability'' would beg the question, and cause 
confusion, since we want to \textit{prove} that it equals the probability.}. 
Since these quantities occur as the diagonal elements of density matrices, we 
can refer to them as ``Born densities'', or, simply, as densities.

With this terminology, and given the motivations described above, wave function 
collapse can be connected to elementary entangling interactions between pre-existing 
particles by the following incremental density shift hypothesis. Consider the 
binary splitting of the wave function of the combined system into interacting 
and noninteracting components. Label the Born density of the interacting component 
as $p$; label the density of the complementary component as $q = 1-p$. Associated 
with each individual entangling interaction of an elementary particle, there is 
a small nonlocal shift, $d$, in the quantities $p$ and $q$, such that 
$p \rightarrow p+d$ and $q \rightarrow q-d$, OR $p \rightarrow p-d$ and 
$q \rightarrow q+d$. The shift is unbiased; that is, the probability of an 
increase in $p$ (and corresponding decrease in $q$) is equal to the probability of 
a decrease in $p$ (and corresponding increase in $q$).\footnote{This hypothesis is 
somewhat similar to the account proposed recently by Bedingham\cite{Bedingham}. It 
is also essentially identical to one made in an earlier unpublished work by this 
author\cite{Gillis}.} The linearity of the ordinary deterministic equations 
permits such amplitude redistribution without disturbing relative phases or 
producing other changes which could enable superluminal signaling. 

The size of the density shifts, $d$, is assumed to be some small number. It 
cannot exceed either $p$ or $q$, since $p$ and $q$ are positive numbers bounded 
by $0$ and $1$. Because this is a phenomenological account, constructed to 
reproduce the macroscopic postulates, there are few other constraints that can 
be placed on it here. In particular, it might vary from one interaction to the 
next. An estimate of the typical size of the shifts will be given in the next 
section based on indirect arguments. 

The hypothetical small, nonlocal, nondeterministic effects are in addition to 
the standard effects of such interactions, which are described within 
conventional quantum mechanics (or quantum field theory). The overall collapse 
process consists of numerous small random density shifts between branches of 
the entangled system. Since an entangled system increases in size as the 
interactions proceed, the scope of the density shifts increases to include more 
and more particles in the collapse process. Eventually the scope can approach 
mesoscopic scales. The expansion of the entangled system also allows many additional 
particles to participate in the interactions (and density shifts). The particular 
way in which the system is split can change from one interaction to the next, 
since different interactions can define different splittings of the wave function. 
This makes possible measurement processes with arbitrarily many possible outcomes. 
As a result of the random density shifts or of the experimental arrangement, one 
particular decomposition eventually becomes dominant, and all amplitude 
is shifted to one of the branches in that pair. For any specific binary 
decomposition, a sequence of steps toward one or the other component has the 
character of a random walk (with possibly varying step size). If $ \overline{d}$ 
is the average step size, then a typical walk will terminate in about 
$(1/\overline{d})^2$ steps. So the hypothesis reproduces the Projection 
Postulate in a very straightforward way. 

Since the density shift proposal was constructed specifically to reproduce 
the Born Rule, showing that it does so is fairly simple. There are three types 
of cases to consider. The simplest situations are those in which the wave 
function is divided into two principal branches, and the particle is detected 
in one of them. The decomposition defined by the interactions remains the same 
throughout the collapse process.\footnote 
     {By "remains the same" I am referring to the decomposition of the initial 
     entangled system; as entangling interactions occur the number of particles 
     correlated with the initial branches of the wave function increases.} 
More general measurement situations allow for different ways of splitting the 
wave function at various stages of the process. The second category covers 
experiments with more than two chains of detector particles, and hence more 
than two possible mutually orthogonal outcomes. The third type includes 
Bell-EPR\cite{Bell_1,Bell_2,EPR} experiments (and their generalizations) 
in which different (measurement) bases can be chosen in spacelike-separated 
regions. The distinctive feature of this third type of case is that the 
states resulting from the measurements can be distinct without being 
completely orthogonal.

The simplest types of measurements consist of sequences of elementary 
entangling interactions sufficiently long to determine which of two orthogonal 
possibilities is realized. Detectors can be placed in either one or both of 
the branches. The key point for these cases is that the binary decomposition 
of the system defined by the interactions of either of the detectors remains 
the same throughout. Label the two branches as $A$ and $B$. For the given 
decomposition, let $p_0$ designate the initial value of the density of 
branch A. According to the Born Rule the probability that a sequence of 
correlating interactions long enough to constitute a measurement will produce 
this component as the outcome is just $p_0$. We can describe the collapse 
brought about by the measurement as a change in the density from $p_0$ to 1.0, 
or from $p_0$ to 0.

Let $Pr(p_0)$ designate the probability of an A outcome predicted by the shift 
hypothesis in this kind of arrangement. We want to show that $Pr(p_0) = p_0$. 
Note first that the boundary conditions are determined by the requirement that 
the step size, $d$, cannot exceed the smaller of $p$ and $q$. Since the density 
lies between 0 and 1, it is obvious that $Pr(0) = 0$, and $Pr(1.0) = 1.0$. 
Given that there is a $0.5$ probability of a step in either direction, and 
that the step size is the same in either direction we get:
  \begin{equation}\label{4p1}
Pr(p_0) = \left(\frac{1}{2}\right)Pr(p_0+d) + \left(\frac{1}{2}\right)Pr(p_0-d)
  \end{equation}  
     (for any $d: 0 \le d \le p_0, 0\le d \le q_0)$.
This expression represents an infinite set of difference equations. One solution 
is $Pr(p_0) = p_0$\cite{Rozanov}. This clearly satisfies both the boundary 
conditions and Eqn. \ref{4p1}.\footnote{An essentially equivalent discussion 
is given in \cite{Gillis}; a similar one is given in \cite{Pearle_1}.} It is 
straightforward to show that the solution is unique.  So, for a sequence of 
elementary interactions in which the binary decomposition defined by those 
interactions does not change, the shift hypothesis yields the Born 
Probability Rule.

The second category includes standard many-outcome (i.e., more than two) experiments. 
The fact that the possible outcomes are all mutually orthogonal keeps the analysis 
simple. For purposes of illustration we can imagine that there are several 
well-defined branches of the wave function, and that there is a detector in each 
branch. With this kind of set-up, at each elementary stage, an interaction from 
one particular detector is selected (as described in Section 2). The branch 
corresponding to this detector will constitute one of the principal components 
of the binary decomposition defined at this stage; we can label this component 
as $O_i$. The complementary component can be designated $O'$; it consists of all 
branches $O_j$ with $j \ne i$. Let us look at one of these other branches, $O_k$. 
Assume that the density of $O_i = p$, that of $O' = q\; (=1-p)$, and that of 
$O_k = r$. The incremental shifts associated with the elementary interactions will 
change the densities of $O_i$ and $O'$ to either $p+d$ and $q-d$, or $p-d$ and $q+d$. 
Since $O_k$ is completely contained in $O'$, we know that $r \le q$. The transfer of 
density between the two orthogonal components of the wave function defined by the 
interaction must leave all phase relationships within each principal component 
undisturbed. This implies that all relative densities within a main branch will 
remain the same for that interaction. Hence the change in the overall density of 
a major component is distributed proportionately among its sub-branches.  Therefore, 
the shift in $r$ will be $(r/q)d = d'$. Since $d$ is the same whether $q$ is 
increased or decreased, $d'$ must also be the same whether it constitutes an 
increase or a decrease. So from the point of view of the $O_k$ component, this shift 
can be viewed as just part of its own random walk with a different step size. Since 
we have explicitly allowed for varying step sizes in the course of the walk, the 
probability that a complete measurement sequence will result in $O_k$  is the 
same as that demonstrated above. So the Born Rule also holds for these kinds 
of arrangements.

The third category involves Bell-EPR type experiments and their generalizations. 
In these situations the binary decompositions of the wave function defined by 
spacelike-separated interactions result in branches that overlap without being 
identical. This means that the possible outcomes of the measurements might be 
neither identical nor orthogonal.  To establish the Born Rule for these cases 
one must show that steps in a random walk in one binary decomposition do not 
bias possible outcomes in a different, overlapping decomposition. I will do this 
by showing that no matter what decomposition is considered, the densities of the 
components in that basis undergo the same kind of unbiased random walk as that 
described above. 

Typical Bell-EPR experiments involve two-particle systems with two states each. 
If $\alpha = -\beta = 1/\sqrt 2$, and if $|x\rangle$ and $|y\rangle$  represent 
up and down spin states, then a two-particle singlet state can be represented as: 
      \[  \alpha|x\rangle|y\rangle + \beta|y\rangle|x\rangle.   \]
With an alternate measurement basis for the second particle, 
\[	|u\rangle = \gamma|x\rangle + \delta|y\rangle, \quad |v\rangle = \delta^*|x\rangle - \gamma^*|y\rangle,  \]
we get:
\[  \begin{array}{ll}  & \alpha|x\rangle|y\rangle + \beta|y\rangle|x\rangle \\

    =\; & \alpha|x\rangle\otimes(\delta^*|u\rangle - \gamma|v\rangle)  + 
\beta|y\rangle\otimes(\gamma^*|u\rangle + \delta|v\rangle) \\

    =\; & \alpha \delta^*|x\rangle|u\rangle - \alpha \gamma|x\rangle|v\rangle + \beta \gamma^*|y\rangle|u\rangle + \beta \delta|y\rangle|v\rangle.
\end{array} \]
						
Suppose that a measurement interaction involving the first particle results 
in a step of size $d$ toward the $|x\rangle$ state. The density of this state 
increases from  $\alpha\alpha^*$ to  $\alpha\alpha^* + d$; the density of 
the $|y\rangle$ state for the first particle decreases from $\beta\beta^*$ 
to $\beta\beta^* - d$. As a result of this shift, the density of the $|u\rangle$ 
state for the second particle changes from  
$  \alpha\alpha^*\delta\delta^* + \beta\beta^*\gamma\gamma^* $ 
to  $(\alpha\alpha^*+d)\delta\delta^* + (\beta\beta^*-d)\gamma\gamma^*$. 
The net change for $|u\rangle$ is $d' = d(\delta\delta^* \, - \, \gamma\gamma^*)$. 
The density of the $|v\rangle$ state changes from
$ \alpha\alpha^*\gamma\gamma^*  + \beta\beta^*\delta\delta^* $ 
to $(\alpha\alpha^*+d)\gamma\gamma^* + (\beta\beta^*-d)\delta\delta^* $, 
a net change of  $d(\gamma\gamma^* \, - \, \delta\delta^*) = -d' $.

So the densities for the orthogonal $|u\rangle|v\rangle$ states change by the 
same amount in opposite directions. If the interaction of the first particle 
had produced a step of size $d$ toward the $|y\rangle$ state, we would have 
gotten an induced step in the $|u\rangle|v\rangle$ basis of identical size but 
with the opposite sign. Again, we can view this as a step in an unbiased random 
walk in the $|u\rangle|v\rangle$ basis, so the derivation of the Born Rule given 
above still applies. It is possible to define more general entangled states. 
Extension of the proof to these situations is straightforward. 

By looking at each elementary entangling interaction as defining a two-way 
splitting of the system, and allowing for equally probable shifts of density 
between the two principal components, we have seen that we can view any 
component in any basis whatsoever as participating in a random walk in which 
its density, $p$, moves between 0 and 1. Its probability of reaching 1.0 is 
just $p$ (if the "walk" is sufficiently long). Thus, the density shift hypothesis 
entails the Born Rule in all simple measurements and in all cases in which 
multiple, commuting measurements take place, even when different bases 
(corresponding to different orthogonal decompositions) are selected 
by detectors in spacelike-separated regions. So the density shift hypothesis 
\textit{reproduces the predictions of standard quantum theory in all 
situations in which questions of causality arise}. 

Conservation laws are maintained in the same way as in conventional theory. 
They are enforced exactly and locally by the dynamic equation governing 
Schr\"{o}dinger evolution. When measurementlike interactions take place, the 
nonlocal density shifts conserve the relevant quantities in a probabilistic 
sense, i.e., the expectation value of a change in the quantity is 
zero.\footnote{In situations such as certain types of scattering experiments, 
additional constraints can insure exact conservation of some quantities.}

The discussion has centered on ``measurements'', but this does not mean that 
collapse only occurs in laboratories when measurements are made. Under the 
density shift hypothesis, collapse does not require a specially designed 
detection apparatus, or the intervention of an intelligent observer. Elementary 
entangling interactions take place all the time, everywhere. Whenever the 
wave function of a particle splits into several branches, or becomes sufficiently 
spread out, so that an interaction with another elementary particle does not 
engage the entire wave function, entanglement results, and density shifts occur. 
A sufficient number of such shifts produce complete state vector 
reduction.\footnote{The view that the registration of measurement results is 
an \textit{objective} process, involving ordinary physical interactions, and 
not dependent on the presence of an intelligent observer, has been expressed 
by many. Notable among these is Bohr, who mentions irreversible acts of 
amplification\cite{Bohr_2}. But it is important to emphasize that the view 
that I am advocating is that the elementary interactions are also perfectly 
objective, though not deterministic, events. This latter view is not shared 
by all.}

In summary, it is possible to construct a microphysical explanation 
of nonlocal measurement effects by attributing those effects to 
elementary correlating interactions.  By assuming that the processes 
that induce wave function collapse are the same ones that observe it, 
we help to preserve causality. By focusing on the interactions that 
constitute measurements we see that they both generate entanglement 
and eventually break it. They also define a natural bifurcation of 
the wave function that explains how the measurement basis is selected; 
this bifurcation suggests a simple density shift hypothesis. If we are 
willing to acknowledge that the density shifts that result in state 
vector reduction are essentially nondeterministic, then we have 
a complete explanation of how causality follows from the fundamental 
laws of physics, and why information can only be realized 
at mesoscopic and larger scales. The in-principle unobservability 
of sequences of spacelike-separated interactions also follows 
as an immediate consequence. It is obvious that \textit{given any 
measurement result and any sequence of shifts}, we can assign 
directions to the shifts that yield the observed result. Since 
observations are inherently mesoscale or larger processes, they 
must be consistent with any sequence of spacelike-separated 
elementary interactions. This is how relativity is 
preserved.

\section{Estimating the Collapse Scale}

In order to design experiments and assess their feasibility we need some 
idea of the size of the deviations from linear evolution hypothesized in 
the previous section. Because this is a phenomenological account we cannot 
derive this from first principles, but we can make an indirect estimate. 
The density shift hypothesis states that the size, $d$,  is less than or 
equal to the density of the interacting component, $p$, and the complementary 
density, $q = 1-p$. Quantum eraser experiments and quantum computer 
implementations involving a few elementary systems have shown that superposition 
effects persist after small numbers of entangling 
interactions\cite{Scully_2,Walborn,Vandersypen,Steffen}. So any deviations from 
perfect linearity must be fairly small compared to the typical amplitudes of the 
particles involved in these experiments. (This point will be further elaborated 
in Section 5.)

Estimates of the collapse scale have been given in previous works that 
treat nonlocal projection effects as real physical phenomena. The dynamical 
reduction theory of Ghirardi, Rimini, and Weber (GRW)\cite{GRW,Ghirardi_Bassi}, 
and the works of Pearle\cite{Pearle_1,Pearle_2} suggest tentative values for 
the relevant parameters. The collapse ``mechanism'' in those accounts is 
different from what has been proposed here, but a comparison can still be made 
in terms of the typical size that an entangled system can reach before 
superposition breaks down.

The dynamical reduction proposals employ a parameter, $\lambda$, which 
describes the frequency with which a typical one-particle wave function 
collapses. For purposes of comparison, we can use the value quoted by 
Pearle\cite{Pearle_1}\footnote{Attributed to GRW\cite{GRW}} of $10^{-16}$ 
per second. To achieve collapse in a relatively small fraction of a second 
requires an entangled system consisting of about  $10^{18}$ (or more) 
elementary particles. This fairly conservative estimate is chosen, in part, 
to minimize the deviations from linear evolution, since no violations of 
the superposition principle have been observed at the level of elementary 
interactions. 

For the hypothesis outlined in Section 3, consistency with current experimental 
results confirming superposition is maintained to a very high degree by the 
standard decoherence mechanism\cite{Zurek1,Zurek_2}. The correlating 
interactions that are assumed to be responsible for nonlocal collapse effects 
make it difficult to observe any deviations from perfect linearity. Since the 
possible deviations tend to be hidden, it is conceivable that the collapse 
scale is much closer to the level of elementary interactions than is assumed 
in most dynamical reduction models. (The next section will explore ways to get 
around this ``masking'' effect, in order to derive feasible experimental 
predictions from the proposals made here.) 

In fact, the general line of argument used in developing this proposal suggests 
a way to estimate the scale of wave function collapse. Information can only be 
instantiated in mesoscale and larger systems because of the probabilistic nature 
of elementary correlating interactions. This means that there is a lower bound on 
the size of naturally occurring information processing systems. We do not need to 
define the characteristics of these systems in precise detail in order to look at 
some examples of them. 

So let us examine some very small biological systems. Given the span of time 
over which they have been evolving, it is reasonable to suppose that nature has 
explored the inherent limits involved in the construction of these entities. We 
will consider only those systems that are capable of operating independently 
within their environments, and carrying out all biological functions, including 
reproduction. This reasonably conservative assumption excludes viruses. 

These organisms must be capable of exchanging energy with their environment to 
develop and maintain their internal organization. Some of their interactions, 
such as absorbing photons, involve probabilistic quantum processes. To enhance 
their prospects for survival and reproduction, it is plausible to suppose that 
these systems must be capable of independently resolving these processes. In 
other words, they must be able, independently, to completely collapse the wave 
function of the particles with which they interact. 

In the previous section I hypothesized that the collapse process has the character 
of a random walk which terminates when the density of one of the branches reaches 
either $0$ or $1$. If the bifurcation defined by the interactions remains the same 
throughout the process, then this takes about $(1/\overline{d})^2$ steps, where 
$ \overline{d}$ is the average step size for the type of interaction involved. 
So organisms capable of inducing collapse, and remaining viable against the 
probabilistic outcomes of the collapse process must consist of somewhat more 
than $(1/\overline{d})^2$ elementary systems. It seems reasonable to suppose 
that the smallest such successful organisms must be on the order of $10$ to 
$100$ times the size of a system minimally capable of bringing about collapse. 

The smallest microbes have linear dimensions of a few hundred nanometers\cite{Lemcke}. 
This sets a rough size limit of about $10^{10}$ to $10^{11}$ elementary particles. 
If the organism is $10$ to $100$ times the size of an average collapse mechanism, 
we get an upper limit for an average wave function collapse of about $10^8$ to 
$10^{10}$ elementary interactions. I will take the geometric mean of these numbers, 
$10^9$, as a nominal value for subsequent discussion.

This number represents the typical number of steps in the random walk that leads 
to complete wave function collapse. The average density shift, $\overline{d}$, 
should be roughly equal to the inverse of the square root of this number. So let 
us tentatively take $\overline{d} = 3*10^{-5}$, as our estimate of the average 
shift involved in ordinary elementary interactions.\ This will be used in the 
next section which outlines some experimental approaches to determine whether, 
and at what scale collapse occurs.

\section{Experimental Consequences}

Wave function collapse is not consistent with strictly linear Schr\"{o}dinger 
evolution. Collapse entails the elimination of superposition effects at some 
stage of the measurement process. However, for a long time after the development 
of quantum theory, there was a tendency to dismiss or ignore the possibility 
of determining at what stage of the process the deviation from linear evolution 
occurs. This tendency was due, in large part, to the influence of Bohr\cite{Bohr_1} 
and Heisenberg\cite{Heisenberg_1,Heisenberg_2}, who argued that it makes no 
difference where one draws the line between the measured system and the measuring 
apparatus.

The plausibility of the Bohr-Heisenberg argument is based on the difficulty of 
observing superposition effects in a system that has undergone interactions. 
The wave functions of single particles that have not interacted can be recombined 
fairly easily, and superposition can be exhibited through straightforward 
interference phenomena, as in simple double-slit or Stern-Gerlach experiments. 
If interactions occur, the branches of the wave function separate more widely in 
configuration space. Recombining the branches requires detailed control of all 
of the particles represented by the entangled wave function. Such detailed control 
is hard to achieve, so in most cases, the particles undergo further interactions. 
Branches of the wave function separate to an even greater extent, and any 
practical possibility of restoring coherence is lost\cite{Zurek1}. In such 
a chain of interactions, the boundary between "system" and "apparatus" is 
clearly somewhat arbitrary.  

Despite these difficulties, there are clear differences in experimental 
predictions between the assumption that the wave function collapses, and the 
view that unmodified linear evolution continues. These differences persist, 
in principle, to any level of entanglement generated by a chain of interactions. 
This point was emphasized by Bell\cite{Bell4}.\footnote{For a somewhat different 
viewpoint, see Janssens and Maassen\cite{Janssens}.}

The first practical tests of the persistence of superposition effects after 
interactions occur were proposed by Scully and his colleagues in the 
1980's\cite{Scully_Druhl,Scully_ES_1,Scully_ES_2}. The key idea in these 
quantum eraser experiments is to separate the wave function of an elementary 
system into two (or more) branches, and have these branches (or at least one 
of them) interact with another "target" elementary system. One then recombines 
the branches of both subject and target, measures them in an alternate basis, 
and looks for correlations between them.  

To illustrate the idea, consider an electron in a z-up state. If the x-up 
and x-down components are separated in a Stern-Gerlach device and then 
recombined without interacting with any other systems, then the superposition 
principle allows us to reconstruct the z-up state and detect it with 
a probability of 1.0 (assuming appropriate phases are maintained, with 
$\alpha = \beta = 1/\sqrt 2$):    
\begin{equation}\label{x3p4}
|z\!\uparrow\rangle \; \Longrightarrow \; \alpha|x\!\uparrow\rangle + \beta|x\!\downarrow\rangle \; \Longrightarrow \; |z\!\uparrow\rangle.
\end{equation}
However, if at least one of the x-spin branches is allowed to interact with 
another elementary system prior to recombination, the possibility for simple 
interference is eliminated, and the probability of a z-up detection is reduced 
to 0.5: 
($\gamma = \delta = 1/\sqrt 2$)
 \begin{equation}\label{eqno1}
\begin{array}{ll} 
	|z_1\!\uparrow\rangle \; \Rightarrow \; \alpha|x_1\!\uparrow\rangle + \beta|x_1\!\downarrow\rangle \;& \Longrightarrow \;                      
	\alpha|x_1\!\uparrow\rangle|x_2\!\uparrow\rangle + \beta|x_1\!\downarrow\rangle|x_2\!\downarrow\rangle \\
    &= \; \gamma|z_1\!\uparrow\rangle|z_2\!\uparrow\rangle + \delta|z_1\!\downarrow\rangle|z_2\!\downarrow\rangle.
\end{array}
\end{equation}   
(In the experiment proposed by Scully, Englert, and Schwinger\cite{Scully_ES_1} 
the role of the target is played by a pair of micromaser cavities. In these 
expressions  $|x_2\!\uparrow\rangle,|x_2\!\downarrow\rangle$ correspond to 
micromaser number states, and $|z_2\!\uparrow\rangle,|z_2\!\downarrow\rangle$ 
represent micromaser symmetric/antisymmetric states. )

Despite the fact that we cannot reconstruct the z-up state in every case, 
superposition is still exhibited in the perfect correlations between detections 
of $|z_1\!\uparrow\rangle$ and $|z_2\!\uparrow\rangle$ (a z-up electron state and 
micromaser symmetric states), and also between $|z_1\!\downarrow\rangle$ and 
$|z_2\!\downarrow\rangle$. The dependence of the correlations on superposition 
can be shown explicitly by expanding the terms in equation \ref{eqno1} separately: 
 \begin{equation} 
\begin{array}{lcl} \alpha|x_1\!\uparrow\rangle|x_2\!\uparrow\rangle 
     & =  &\alpha(1/\sqrt{2})^2(|z_1\!\uparrow\rangle + |z_1\!\downarrow\rangle)\otimes(|z_2\!\uparrow\rangle +  |z_2\!\downarrow\rangle) \\
     & =  &\alpha(1/\sqrt{2})^2(|z_1\!\uparrow\rangle|z_2\!\uparrow\rangle + |z_1\!\downarrow\rangle|z_2\!\downarrow\rangle 
                        + |z_1\!\uparrow\rangle|z_2\!\downarrow\rangle + |z_1\!\downarrow\rangle|z_2\!\uparrow\rangle) 
  \end{array}
 \label{eqno2}
 \end{equation}
and 
 \begin{equation}\label{eqno3}
\begin{array}{lcl} \beta|x_1\!\downarrow\rangle|x_2\!\downarrow\rangle 
     & =  &\beta(1/\sqrt{2})^2(|z_1\!\uparrow\rangle - |z_1\!\downarrow\rangle)\otimes(|z_2\!\uparrow\rangle - 
 |z_2\!\downarrow\rangle) \\
     & =  &\beta(1/\sqrt{2})^2(|z_1\!\uparrow\rangle|z_2\!\uparrow\rangle + |z_1\!\downarrow\rangle|z_2\!\downarrow\rangle 
                         - |z_1\!\uparrow\rangle|z_2\!\downarrow\rangle - |z_1\!\downarrow\rangle|z_2\!\uparrow\rangle)
\end{array}
\end{equation}  
The form in \ref{eqno1}, which displays the perfect correlations, results because 
of the cancellation of the up-down cross terms, 
($|z_1\!\uparrow\rangle|z_2\!\downarrow\rangle$ 
and  $|z_1\!\downarrow\rangle|z_2\!\uparrow\rangle$), that occurs 
when \ref{eqno2} and \ref{eqno3} are superposed (since $\alpha = \beta$). 

We can see the clear inconsistency between the assumptions of continued 
linear evolution and projection by supposing that the interaction between the 
electron and the micromaser cavity had constituted a full measurement (in the 
sense of inducing complete collapse). In that case the coefficients, $\alpha$ 
and $\beta$, would have been changed to 1 and 0, or 0 and 1. The correlations 
would be totally eliminated. 

This inconsistency is \textit{not} removed by adding more interactions. To see 
this, consider an  entangled system with many particles. Label the states of the 
"subject" and (multiple) "detector" systems as  		
$|x_s\rangle,|z_s\rangle, |x_{di}\rangle, |z_{di}\rangle$. Designate the 
superposition of all normalized product states with an even number of 
$|z\!\downarrow\rangle$ detector states as $|z_d\!\downarrow_{EVEN}\rangle$, 
with $|z_d\!\downarrow_{ODD}\rangle$ representing the complementary 
states.\footnote{Even numbers include zero, which covers the case described 
by \ref{eqno1}.} (Normalization factor $ = (1/\sqrt{2})^{N-1}. $) After $N$ 
correlating interactions in the x-spin basis, we get
 \begin{equation}\label{3p8}
 \begin{array}{lcl}
    &   &  \alpha|x_s\!\uparrow\rangle|x_{d1}\!\uparrow\rangle...|x_{dN}\!\uparrow\rangle \\
    & = &  \alpha(1/\sqrt{2})^{N+1}[\,(|z_s\!\uparrow\rangle \, +  \, 
|z_s\!\downarrow\rangle)\otimes(|z_{d1}\!\uparrow\rangle \, + \, |z_{d1}\!\downarrow\rangle)...(|z_{dN}\!\uparrow\rangle \,+ \, |z_{dN}\!\downarrow\rangle)\,] \\
    & = &  (\alpha/2)[\,|z_s\!\uparrow\rangle|z_d\!\downarrow_{EVEN}\rangle \, + \, |z_s\!\downarrow\rangle|z_d\!\downarrow_{ODD}\rangle  \, + \, |z_s\!\uparrow\rangle|z_d\!\downarrow_{ODD}\rangle \, + \, |z_s\!\downarrow\rangle|z_d\!\downarrow_{EVEN}\rangle ],  
  \end{array}   
\end{equation}   
and 
 \begin{equation}\label{3p9}
 \begin{array}{lcl}
   &   &  \beta|x_s\!\downarrow\rangle(|x_{d1}\!\downarrow\rangle...|x_{dN}\!\downarrow\rangle  \\
   & = &  \beta(1/\sqrt{2})^{N+1}[\,(|z_s\!\uparrow\rangle \, - \,  |z_s\!\downarrow\rangle)\otimes(|z_{d1}\!\uparrow\rangle \, - \, |z_{d1}\!\downarrow\rangle)...(|z_{dN}\!\uparrow\rangle \, - \, |z_{dN}\!\downarrow\rangle] \\
   & = & (\beta/2)[\,|z_s\!\uparrow\rangle|z_d\!\downarrow_{EVEN}\rangle \, + \, |z_s\!\downarrow\rangle|z_d\!\downarrow_{ODD}\rangle 
 \, - \, |z_s\!\uparrow\rangle|z_d\!\downarrow_{ODD}\rangle) \, - \, |z_s\!\downarrow\rangle|z_d\!\downarrow_{EVEN}\rangle ].      	 
  \end{array}   
\end{equation}    

If no collapse occurs after $N$ correlating interactions, then $\alpha = \beta$, 
and superposition of the two expressions yields cancellation of the terms,  
$|z_s\!\uparrow\rangle|z_d\!\downarrow_{ODD}\rangle)$ 
and $|z_s\!\downarrow\rangle|z_d\!\downarrow_{EVEN}\rangle$. This gives a perfect 
correlation between up results of z-spin measurements on the subject particle, and 
an \textit{even} number of down results from z-spin measurements on the detector 
particles (with the opposite correlation for down results on the subject 
particle).\footnote{Of course, the detector states can be any appropriate correlated 
states (such as micromaser states); the z-spin designation is just for convenience.} 
If, on the other hand, $\alpha$  and $\beta$ have been changed to 1 and 0, or 0 
and 1 by a collapse, then there is no cancellation of the cross terms, and the 
correlations are eliminated.\footnote{The question of whether and when deviations 
of this general nature occur must be faced by any account of the measurement 
problem. No-collapse interpretations such as that of Everett\cite{Everett}, 
decoherence accounts\cite{Decohere}, and de Broglie-Bohm theory\cite{deBroglie,Bohm} 
(with ``correct'' initial distributions) predict zero deviations from the perfect 
correlations no matter how large $N$ is. Spontaneous collapse 
theories\cite{GRW,Ghirardi_Bassi,Pearle_1,Pearle_2} imply deviations of some 
magnitude at some stage of the measurement process.} 

It should be noted that the strategy of inducing particles to interact in one 
basis, and measuring them in a different basis constitutes the essential ``trick'' 
of quantum computing, as emphasized by Mermin\cite{Mermin}. So efforts to 
construct quantum computers also serve as tests of whether and when collapse 
occurs. Most of the work in quantum computing proceeds on the assumption that 
one is free to set the boundary between linear Schr\"{o}dinger evolution and 
measurement at any convenient point (aside from pragmatic concerns about 
decoherence). However, the (assumed) linear evolution typically includes correlating 
interactions of the same general sort as are involved in measurement processes. 
If wave function collapse is a real physical process associated with these 
interactions, then it will set in-principle limits on how far out one can place 
the measurement boundary.

Given these considerations, we can summarize the status of current experimental 
results. Expected superposition effects have been observed in double-slit quantum 
eraser experiments\cite{Scully_2}, and quantum computations involving a few 
elementary interactions have been successfully 
implemented\cite{Walborn,Vandersypen,Steffen}. These results indicate that any 
deviations from perfectly linear evolution in individual interactions are small 
compared to the typical amplitudes of the particles involved. 

To assess the feasibility of detecting the deviations that were hypothesized 
in Section 3, let us take another look at the situation represented by equations 
\ref{3p8} and \ref{3p9} above. As stated, if no deviations from linearity 
have occurred after $N$ interactions, then $\alpha = \beta$, and the perfect 
correlation between z-up results on the subject particle, and an even number of 
down results on the detector particles is maintained. If, on the other hand, 
density shifts associated with the interactions have changed $\alpha$ and $\beta$ 
to $\alpha'$ and $\beta'$ with $\alpha' \ne \beta'$ then we do not get (complete) 
cancellation, and the correlations are altered. 

If we use the estimate of the average step size, $d$, from the previous section, 
we can make a provisional calculation of the size of the deviations predicted 
by the shift hypothesis for this particular situation. In a random walk of $N$ 
steps the average deviation from the starting point is $\sqrt N$. The initial 
densities are the squared amplitudes, $\alpha\alpha^*$ and $\beta\beta^*$. 
After $N$ interactions these would be altered on average to 
$\alpha'\alpha'^* = \alpha\alpha^* \pm (\sqrt N)d$, and 
$\beta'\beta'^* = \beta\beta^* \mp (\sqrt N)d$. For values of  
$(\sqrt N)d~\ll~\alpha\alpha^*,~\beta\beta^*$, these differences imply changes 
in the amplitudes of approximately $(\sqrt{N})(d/2\alpha)$ and 
$(\sqrt{N})(d/2\beta)$. By substituting the values  $(\alpha \pm d\sqrt{N}/2\alpha)$  
and  $(\beta \mp d\sqrt{N}/2\beta)$  for $\alpha$ and $\beta$ in expressions 
\ref{3p8} and  \ref{3p9} and superposing them, we can compute the amplitude 
coefficients of the cross terms, $(|z_s\!\uparrow\rangle|z_d\!\downarrow_{ODD}\rangle)$ 
and $(|z_s\!\downarrow\rangle |z_d\!\downarrow_{EVEN}\rangle)]$.  If the initial 
values of $\alpha$ and $\beta$ are  $1/\sqrt{2}$, then the amplitude for the 
cross terms is approximately $(\sqrt{N/2})d$. So the probability of detecting 
these "deviant" cases is about $(N/2)d^2$. The estimate for $d^2$ from Section 4 
was of order  $10^{-9}$. It would obviously be quite difficult to detect such 
deviations in these kinds of experimental arrangements. 

To increase the likelihood of observing possible departures from linear evolution, 
we need to maximize the deviations in the \textit{amplitudes} of the components. 
Our working assumption is that deviations in the \textit{densities} will be roughly 
the same size in most cases.\footnote{The reason for framing the hypothesis in 
terms of density shifts instead of amplitude shifts was to maintain strict adherence 
to the Born Rule, and thus preserve causality.} The small size of the value  derived 
above, $(\sqrt{N/2})d$, results largely from the condition, 
$(\sqrt N)d~\ll~\alpha\alpha^*,~\beta\beta^*$. To increase the size of the possible 
amplitude shifts, we can work with states in which $d~\approx~\alpha\alpha^*$. 
This will give us amplitude differences of order $\sqrt{d}$, instead of $d$. 

We can calculate the probability of a deviation from linear evolution in terms 
of $d$ as follows. As before, we begin with a state, 
$\alpha|x_s\!\uparrow\rangle + \beta|x_s\!\downarrow\rangle$, but now 
with $\alpha \ll \beta$. Entanglement is generated by inducing one of the x-spin 
branches to interact with a detector system; we then look for correlations with 
z-spin branches:
 \[   (\alpha|x_s\!\uparrow\rangle + \beta|x_s\!\downarrow\rangle) \otimes |x_d-ready\rangle 
   \Longrightarrow \; \alpha|x_s\!\uparrow\rangle|x_d\!\uparrow\rangle \, + \, \beta|x_s\!\downarrow\rangle|x_d\!\downarrow\rangle\] 
If $\alpha \ll \beta$, the resulting state can still be represented in the form: 
 \[   (1/\sqrt{2})|z_s\!\uparrow\rangle|u_d\rangle \, + \, (1/\sqrt{2})|z_s\!\downarrow\rangle|v_d\rangle, \]   
but now the correlates of $|z_s\!\uparrow\rangle$ and $|z_s\!\downarrow\rangle$ 
are not orthogonal. Assuming no collapse effects they can be easily calculated: 
 \[  |u_d\rangle \, = \, (\alpha+\beta)(1/\sqrt{2})|z_d\!\uparrow\rangle \, + \, (\alpha-\beta)(1/\sqrt{2})|z_d\!\downarrow\rangle, \] 
 \[  |v_d\rangle \, = \, (\alpha-\beta)(1/\sqrt{2})|z_d\!\uparrow\rangle \, + \, (\alpha+\beta)(1/\sqrt{2})|z_d\!\downarrow\rangle. \] 
With no deviation from linearity one should still observe perfect correlations 
between $|z_s\!\uparrow\rangle$ and $|u_d\rangle$ and between 
$|z_s\!\downarrow\rangle$ and $|v_d\rangle$.  However, if there is a shift, $d$, 
in the density due to the interaction, then $\alpha$ and $\beta$ are shifted to 
$\alpha'$ and $\beta'$, and $|u_d\rangle$ and $|v_d\rangle$ are shifted to 

  \[ |u'_d\rangle = (\alpha'+\beta')(1/\sqrt 2) |z_d\!\uparrow\rangle + (\alpha'-\beta')(1/\sqrt 2) |z_d\!\downarrow\rangle, \] 
  \[ |v'_d\rangle = (\alpha'-\beta')(1/\sqrt 2) |z_d\!\uparrow\rangle + (\alpha'+\beta')(1/\sqrt 2) |z_d\!\downarrow\rangle. \] 

The effect on the correlations between the $|z_s\rangle$ and original 
$|u_d\rangle$ and $|v_d\rangle$ states can be computed by taking the inner 
product between the original and primed states. With $\alpha$ and $\beta$ 
real (for simplicity) we get 
 \[
  \begin{array}{lcl}
     \langle u'_d|u_d\rangle & = & (1/2){(\alpha\alpha' +\beta\beta'+\alpha\beta'+\beta\alpha') + (\alpha\alpha'+\beta\beta'-\alpha\beta'-\beta\alpha')}  \\

     & = & (\alpha\alpha' + \beta\beta') 
  \end{array}
 \]
\[ 
 \begin{array}{lcl}  
  \langle v'_d|v_d\rangle & = & (1/2){(\alpha\alpha'+\beta\beta'-\alpha\beta'-\beta\alpha') + (\alpha\alpha'+\beta\beta'+\alpha\beta'+\beta\alpha')} \\
     & = & (\alpha\alpha'+\beta\beta') 
 \end{array}
 \]                 
Collapse effects (density shifts) are indicated by deviations from the perfect 
correlations predicted by strictly linear evolution. Their magnitude is: 
\[ 1 - (\alpha\alpha' + \beta\beta')^2 = 1 - ( \alpha\alpha\alpha'\alpha' + \beta\beta\beta'\beta' 
+ 2\alpha\alpha'\beta\beta').  \]

Suppose that $\alpha\alpha = d$. The density shift hypothesis implies that 
in half the cases $\alpha\alpha$ would be shifted to $\alpha'\alpha'= 0$, 
and $\beta\beta$ would be shifted to $\beta'\beta'= 1$. In these situations 
we get $1 - \beta\beta = \alpha\alpha$; the statistical deviation for this half 
of the cases will be  $\alpha\alpha = d$. In the other half of the cases 
$\alpha\alpha$ would be increased to $\alpha\alpha+d = 2\alpha\alpha$, and 
$\beta\beta$ would be decreased to $\beta\beta-d$. 
With $\beta\beta = 1-\alpha\alpha$ and $\beta'\beta'  = 1 - 2\alpha\alpha$, 
$\beta \approx 1 - \alpha\alpha/2, \alpha' = \sqrt{2}\alpha$, we get 
\[      1 - [(1-\alpha\alpha)(1-2\alpha\alpha) + 2\alpha\alpha\alpha\alpha + 2 \sqrt 2 \alpha\alpha(1-\alpha\alpha/2)(1-\alpha\alpha)] \]
\[        = 3\alpha\alpha - 2.8\alpha\alpha - (higher\; order\; terms) \qquad  \approx \qquad   0.2\alpha\alpha.       \]

So in half the cases the deviation is $d$, and in the other half it is about 
$0.2d$, for an average of about $0.6d$. The estimate for $d$ is about $3*10^{-5}$ 
(approximately 1 part in 30,000), so it might be experimentally accessible. 
Of course, with $\alpha$ being small there would be a need to control and measure 
the initial state very precisely. 

Although experimental tests of wave function collapse face substantial challenges, 
the efforts being made to construct quantum computers give some hope that such tests 
can be made in the not too distant future. If quantum computers reach the stage 
where the entangled systems involved in the computations include millions of 
elementary particles, the possible deviations from perfect linearity could become 
noticeable.

In closing this section it is important to emphasize that \textit{the deviations from 
linear evolution predicted by the density shift hypothesis do not violate causality}. 
To observe any deviations (if they exist), one must recombine branches of the wave 
function, and measure them in a basis other than the one in which they have 
interacted. This implies that the final measurements must be \textit{timelike later} than 
those interactions that established the initial entanglement relations, so the 
possibility of superluminal signaling never arises.

\section{Discussion}

The argument of the previous section shows that there are clear inconsistencies 
in the experimental predictions of quantum theory depending on where one draws 
the line between measured system and measuring apparatus. This problem stems 
from the clash between the deterministic, (mostly) local evolution described by 
the dynamic equations governing elementary processes, and the probabilistic, 
nonlocal nature of the measurement postulates. These inconsistencies can be 
obscured by the size and complexity of measurement arrangements, but they are 
always there, in principle. As experimental techniques advance, and quantum 
computational apparatus becomes more sophisticated, the potential conflicts will 
become even sharper. 

To resolve the discrepancies, we must carefully examine the conceptual framework 
of contemporary physics. The nonlocal and nondeterministic effects described by 
the measurement postulates resist easy incorporation into a relativistic spacetime 
structure. Nevertheless, the standard version of quantum measurement does share 
at least one key property with conventional interpretations of relativity based on 
local deterministic theories. Conventional relativity and standard measurement 
theory both prohibit the superluminal transmission of information. Since signal 
causality is the most obvious feature that these two apparently disparate aspects 
of current theory have in common, it makes sense to take this principle as 
a starting point in any attempt to reconcile them. 

As stated in the introduction, Bell criticized this approach because it can be 
both imprecise and subjective. His arguments have merit, but the proper response 
should not be to abandon the only readily identifiable feature shared by the 
"two fundamental pillars of contemporary theory"\cite{Bell3}. Instead, we should 
try to fix the shortcomings. 

We can eliminate the problem of subjectivity by noting the connection of causality 
to the concept of information, and insisting that this concept be definable in 
purely physical terms. We do not need to develop a complete account of what it 
means for physical systems to instantiate and transmit information, but we do 
need to be sure that any inferences that we draw based on the concept are 
consistent with an interpretation of information as a strictly physical 
relationship. 

When we take this approach, we immediately see some very specific consequences 
of signal causality. Given the nonlocality of projection effects, one can 
immediately derive the Born Probability Rule, as demonstrated in Gleason's theorem. 
It also follows that any physical process capable of observing nonlocal effects must 
be capable of inducing those effects, so projection must be tied to measurementlike 
processes. By extending this inference to the microphysical level, we see that 
the probabilistic, nonlocal effects must stem from the elementary interactions that 
generate entanglement.

So, the limits on determinism that are implied by these considerations reconcile 
the existence of real nonlocal effects with the impossibility of using them to 
transmit information. Because the interactions that constitute measurements are 
not deterministic, information can only be instantiated in systems that involve 
enough correlating interactions and replicating subsystems to fully resolve 
probabilistic processes and insure reasonably stable records of the outcomes. 
When states are changed by spacelike-separated interactions, the local physical 
record of the previous state is eliminated, unless it is stored in a system 
capable of recording, interrogating, and comparing the states of a sufficient 
number of subsystems. These types of processes are of the sort that bring about 
wave function collapse. So there is never any physical representation that 
a collapse has been brought about by a spacelike-separated set of interactions. 

The specific consequences and general insights that follow from the principle 
of signal causality, properly interpreted, should give us some confidence that 
it can be further exploited to help bridge the gap between relativity and 
quantum nonlocality.

To resolve the conflict, we must first recognize how sharp it is. Any 
coherent explanation of wave function collapse at the level of elementary 
interactions must deal with two related types of nonlocal effects: 
(1) the familiar ones that project out one particular branch of an 
entangled state, and (2) the generation of the entanglement relations 
that mediate the collapse and define the subsystems (and states) that are 
being acted on. In any realistic collapse process there are huge 
numbers of spacelike-separated elementary interactions that are involved 
in myriad nonlocal effects. A consistent microphysical description of 
projection must be able to deal with this complexity.

We can construct a very straightforward account of probabilistic, nonlocal 
effects by assuming that there is \textit{some sequence} to those effects. 
With such an assumption, the state on which those effects act is well defined, 
and the need for cooperation among the spacelike-separated processes that 
generate the effects is eliminated. Any attempt to describe collapse at the 
elementary level without such a sequencing assumption faces extreme difficulties.

Of course, any objective sequencing of spacelike-separated events is at odds 
with the idea that spacetime structure is completely described by relativity. 
To deal with this dilemma, consider the intimate connection between causality 
and relativity. We have seen that the key to preserving causality in the face 
of nonlocal measurement effects is the nondeterministic nature of those effects. 
If we assume that these two defining properties stem from the same roots, 
then that nondeterministic nature should also explain the emergence of relativity. 

The indeterminism of nonlocal collapse effects preserves signal causality not 
by negating the reality of those effects, but rather by requiring any 
\textit{observational} process to be of macroscopic (or at least mesoscopic) scale. 
The same indeterminism preserves relativity not by denying the objective existence 
of sequences of spacelike-separated interactions, but rather by insuring that those 
sequences lie below the threshold of observability. So relativity is not a complete 
characterization of spacetime structure. It is a property of the family of 
empirically equivalent descriptions of the processes that occur in spacetime. The 
lack of determinism in elementary interactions means that objectively distinct 
sequences of elementary events correspond to the same set of observations. This 
is what guarantees \textit{the equivalence of all empirically definable 
reference frames}.

The overarching lesson is that Bell's analysis has shown that we cannot unify 
physics by explaining macroscopic measurement results in local deterministic 
terms. Instead we must find a way to incorporate the nonlocal, nondeterministic 
aspects of reality into our description of elementary processes. We can do this 
by recognizing the following key points. Signal causality implies that nonlocal 
measurement effects originate in the elementary interactions that generate 
entanglement, and that these effects are essentially probabilistic. A coherent 
account of these effects requires that spacelike-separated interactions involved 
in a collapse process have some objective sequence. The indeterminism that 
protects signal causality also maintains relativity by making these sequences 
unobservable, in principle.

Once these points are granted, it is fairly easy to construct a phenomenological 
description of collapse at the microphysical level. With every elementary 
interaction that generates entanglement we associate a probabilistic shift 
of density (squared amplitude) either into or out of the interacting branch. 
The binary decomposition of the wave function into interacting and noninteracting 
branches provides a natural definition of what is being measured. The Projection 
Postulate and the Born Probability Rule are readily reproduced. Together 
with the recognition that information can only be instantiated at mesoscopic and 
larger scales, this maintains signal causality, and preserves the relativistic 
transformation properties of empirically equivalent descriptions of events 
in spacetime.  

In closing, it is necessary to point out some limitations of the ideas presented 
here. First, this account is just a phenomenological outline of a solution to 
the problem. If it is on the right track, it should eventually be supplanted by 
a more thorough mathematical treatment that would show how the standard quantum 
description in relativistic spacetime can be recovered by averaging over the 
possible sequences of probabilistic nonlocal effects. At a deeper level, there 
remains a sort of dualism in the picture of physical processes and spacetime. 
Most deterministic processes are still confined to the light cone, while certain 
types of nondeterministic effects can propagate across spacelike intervals. The 
concept of signal causality summarizes what is common to these two aspects of 
physical reality, but it does not provide a clear vision of how these apparently 
distinct manifestations emerge from a unifying principle.

Nevertheless, we can move toward a more unified view by better understanding
the connection between macroscopic and microscopic realms. To do this we need to 
integrate the measurement postulates with our theories of elementary interactions
by taking seriously the nonlocality and the limits on determinism implied by 
those postulates.

\section*{Acknowledgements} 

I am grateful to Tim Allison, Brian E. Howard, and Ted Metzler for their 
encouragement, and for reviewing earlier versions of this paper. Thanks to 
Daniel Rohrlich for a detailed account of the insights leading to the 
recognition of the connection between nonsignaling-nonlocal theories and limits 
on determinism. Thanks to Matt Leifer for observations on the ontological status 
of the wave function. I would also like to thank John Taylor for his assistance 
with an earlier related work.

\end{document}